\documentclass[twocolumn,aps,showpacs,preprintnumbers,amsmath,amssymb, superscriptaddress]{revtex4-1}
\pdfoutput=1
\usepackage{bm}
\usepackage{graphicx}
\usepackage{color}
\usepackage{xspace}
\definecolor{dblue}{rgb}{0,0,0.6}
\definecolor{dred}{rgb}{0.9,0,0}
\definecolor{dgreen}{rgb}{0,0.4,0}
\makeatletter

\newcommand{\Rmnum}[1]{\expandafter\@slowromancap\romannumeral #1@}
\makeatother

\begin{document}

\title{Raman scattering investigation of the electron-phonon coupling in superconducting Nd(O,F)BiS$_2$}

\author{S. F. Wu}
\affiliation{Beijing National Laboratory for Condensed Matter Physics, and Institute of Physics, Chinese Academy of Sciences, Beijing 100190, China}
\author{P. Richard}\email{p.richard@iphy.ac.cn}
\affiliation{Beijing National Laboratory for Condensed Matter Physics, and Institute of Physics, Chinese Academy of Sciences, Beijing 100190, China}
\affiliation{Collaborative Innovation Center of Quantum Matter, Beijing, China}
\author{X. B. Wang}
\affiliation{Beijing National Laboratory for Condensed Matter Physics, and Institute of Physics, Chinese Academy of Sciences, Beijing 100190, China}
\author{C. S. Lian}
\affiliation{Beijing National Laboratory for Condensed Matter Physics, and Institute of Physics, Chinese Academy of Sciences, Beijing 100190, China}
\author{S. M. Nie}
\affiliation{Beijing National Laboratory for Condensed Matter Physics, and Institute of Physics, Chinese Academy of Sciences, Beijing 100190, China}
\author{J. T. Wang}
\affiliation{Beijing National Laboratory for Condensed Matter Physics, and Institute of Physics, Chinese Academy of Sciences, Beijing 100190, China}
\author{N. L. Wang}
\affiliation{International Center for Quantum Materials, School of Physics, Peking University, Beijing 100871, China}
\affiliation{Collaborative Innovation Center of Quantum Matter, Beijing, China}
\author{H. Ding}\email{dingh@iphy.ac.cn}
\affiliation{Beijing National Laboratory for Condensed Matter Physics, and Institute of Physics, Chinese Academy of Sciences, Beijing 100190, China}
\affiliation{Collaborative Innovation Center of Quantum Matter, Beijing, China}

\date{\today}

\begin{abstract}
We have performed polarized Raman scattering measurements on the newly discovered superconductor Nd(O,F)BiS$_2$ ($T_c = 4$ K). We observe 2 Raman active  modes, with frequencies in accordance with first-principles calculations. One A$_{1g}$ phonon mode at 112.4 cm$^{-1}$ exhibits a Fano line shape due to electron-phonon coupling. We find a resonance for this mode at 2.45 eV excitation energy. We estimate a 0.68 contribution of this mode to the electron-phonon coupling constant $\lambda$. Our Raman results suggest that the BiS$_2$-based superconductors are possibly phonon-mediated BCS superconductors.
\end{abstract}



\maketitle

Despite low superconducting (SC) critical temperatures ($T_c$'s) of 3 K - 10 K in Bi$_4$O$_4$S$_3$ and (La/Nd/Ce/Pr)O$_{0.5}$F$_{0.5}$BiS$_2$ \cite{Mizuguchi_PRB86,Shiva_JASS134,Yajima_JPSJ81,Masanori_JPSP82,
S.G.Tan_Eur.Phys.J.B85,Awana_Solid201321,K.Deguchi_EPL101,Xing_PRB86,Joe2014,Yoshikazu2014,Selvan_RRL7}, the BiS$_2$-based superconductors are often regarded as possible high-temperature superconductor candidates. Indeed, their square BiS$_2$ layers are quite similar to the CuO$_2$ layers in high-$T_c$ cuprate superconductors \cite{Bednorz1986} and the FeAs planes in the iron-based superconductors \cite{Hosono_JACS130}. First-principles calculations \cite{Wan_PRB87,Li_EPL101} indicate that LaO$_{0.5}$F$_{0.5}$BiS$_{2}$ evolves from a band insulator to a bad metal upon doping. Interestingly, a large electron-phonon coupling constant $\lambda=0.85$ is predicted, leading to a $T_c$ value of 10.6 K from the Allen-Dynes formula \cite{Allen_PRB12,McMillan_PhysRev167} that is compatible with experiments, suggesting a phonon-mediated pairing mechanism. Recent angle-resolved photoemission spectroscopy (ARPES) data were also interpreted in terms of the presence of polarons \cite{Zeng.arXiv1402.1833}. However, a neutron scattering work \cite{Lee_PhysRevB.87} suggests that the electron-phonon coupling in LaO$_{0.5}$F$_{0.5}$BiS$_{2}$ is weaker than expected and a strong SC paring exceeding the limit of simple phonon mediated scenarios was proposed from upper critical field and magnetoresistance measurements \cite{Li_SciChina56}. Consequently, whether phonons are the paring glue for the BiS$_2$-based superconductors is still under debate.

In this paper we use Raman scattering to investigate directly the crystallographic structure and the electron-phonon coupling in Nd(O,F)BiS$_2$ single-crystals. We report 2 Raman active modes, with frequencies in accordance with first-principles calculations. A photon energy resonance at about 2.45 eV for the 112.4 cm$^{-1}$ phonon mode is observed and assigned to an electronic transition. The asymmetric phonon spectral profile of this mode indicates a strong electron-phonon coupling and suggests that phonons are possibly the paring glue in the BiS$_2$-based superconductors.

The Nd(O,F)BiS$_2$ single-crystals used in our Raman scattering measurements were grown by a flux method with KCl/LiCl as the flux \cite{Wang_arxiv2014}. Energy-dispersive X-ray spectroscopy (EDX) measurements performed on several pieces of samples with $T_c=4$ K give an averaged composition of Nd$_{0.95\pm0.02}$O$_{0.56\pm0.1}$F$_{0.44\pm0.1}$Bi$_{0.94\pm0.02}$S$_{2}$ \cite{Wang_arxiv2014}. The crystals were cleaved in air to obtain flat surfaces and then quickly transferred into a low-temperature cryostat ST500 (Janis) for the Raman measurements between 5 and 300 K under a working vacuum better than $2\times 10^{-6}$ mbar. Raman scattering measurements were performed using Argon-Krypton laser lines in a back-scattering micro-Raman configuration with a triple-grating spectrometer (Horiba Jobin Yvon T64000) equipped with a nitrogen-cooled CCD camera. For resonance Raman investigations different ions gas laser lines (1.8 eV - 2.7 eV, 457.9 - 676.4 nm) were used with a constant power of P = 3 mW. Spectra obtained with different laser lines were corrected for the instrumental spectral response. The beam power was reduced to avoid local heating, and was kept below 3 mW during our measurements at low temperature. In this manuscript, we define $x$ and $y$ as the directions along the $a$ or $b$ axes, oriented along the Bi-Bi bounds, while $x'$ and $y'$ are defined at 45$^{\circ}$ from the Bi-Bi bounds. The $z$ direction corresponds to the $c$-axis perpendicular to the BiS$_{2}$ planes.

\begin{table}[!t]
\caption{\label{EXP_LDA}Comparison of the calculated and experimental phonon modes at 294 K.}
\begin{ruledtabular}
\begin{tabular}{cccccc}
 Sym. &	Act.     &	  Exp.   &	      Cal.      & Main atom displacements \\
\hline
E$_u$   &   IR &             &       41.4      & Nd(xy),Bi(-xy)\\
E$_{g}$& 	Raman& 	         &	     43.3	    & Bi(xy),S$^2$(xy)\\
A$_{2u}$& 	IR& 		     &       59.9	    & S$^2$(z)\\
A$_{1g}$& 	Raman& 		 52.6&       70.2 	    & Bi(z)\\
E$_{g}$& 	Raman& 		     &       76.0	    & Nd(xy)\\
E$_{g}$&    Raman& 	         & 	    101.9 	    & S$^2$(xy)\\
A$_{1g}$& 	Raman& 		112.4&      108.3	    & S$^2$(z)\\
A$_{2u}$& 	IR& 		     &      118.1	    & Bi(z),S$^2$(z)\\
E$_{u}$& 	IR& 	         & 	    120.2 	    & S$^2$(xy)\\
E$_{u}$& 	IR& 	         & 	    143.0 	    & S$^1$(xy)\\
E$_{g}$& 	Raman& 	         &      146.6	    & S$^1$(xy)\\
E$_{g}$& 	Raman& 		     &      160.8      & Nd(z)\\
E$_{u}$& 	IR& 	         &      214.1	    & O(xy)\\
B$_{1g}$& 	Raman& 	         &      259.5 	    & O(z)\\
A$_{2u}$& 	IR& 	         &      274.5 	    & O(z)\\
A$_{1g}$& 	Raman& 	         &	    328.5	    & S$^1$(z)\\
E$_{g}$& 	Raman& 		     &      396.3	    & O(xy)\\
A$_{2u}$& 	IR& 		     &      399.2	    & O(z)\\
\end{tabular}
\end{ruledtabular}
\begin{raggedright}
S$^1$ in the Bi-S layer; S$^2$ out of the Bi-S layer\\
\end{raggedright}
\end{table}

The crystal structure of the parent compound NdOBiS$_{2}$ is characterized by the space group D$_{4h}^{7}$(P4/nmm). A simple group symmetry analysis \cite{Comarou_Bilbao} indicates that the phonon modes at the Brillouin zone (BZ) center decompose into [4A$_{1g}$+B$_{1g}$+5E$_{g}$]+[4A$_{2u}$+4E$_{u}$]+[A$_{2u}$+E$_{u}$], where the first, second and third terms represent the Raman-active modes, the infrared (IR)-active modes and the acoustic modes, respectively. To estimate the phonon frequencies, we performed first-principles calculations in the non-magnetic phase of the phonon modes at the BZ center in the framework of the density functional perturbation theory (DFPT) \cite{Baroni_RMP73}, using the experimental lattice parameters $a=b=3.996$ \AA\xspace and $c=13.464$ \AA\xspace for the parent compound \cite{Masanori_JPSP82}, and the Wyckoff positions Nd 2c, Bi 2c, S 2c, O 2a. We used the Vienna \emph{ab initio} simulation package (VASP) \cite{KressePRB54} with the generalized gradient approximation (GGA) of Perdew-Burke-Ernzerhof  \cite{Perdew_PRL77}for the exchange-correlation functions. The projector augmented wave (PAW) \cite{Blochl_PRB50} method was employed to describe the electron-ion interactions. A plane wave cutoff energy of 520 eV was used with a uniform $11\times 11\times 5$ Monkhorst-Pack $k$-point mesh for integrations over the BZ. The frequencies and displacements of the phonon modes were derived from the dynamical matrix generated by the DFPT method \cite{Baroni_RMP73}. The calculated frequencies, the optical activity of the phonon modes and the main atomic displacements of the Raman-active phonon modes are summarized in Table \ref{EXP_LDA}. For the electronic band structure calculations, we performed first-principles calculations of the electronic structure using the full-potential linearized-augmented plane-wave (FP-LAPW) method implemented in WIEN2K package for the real crystal structure of NdO$_{0.5}$F$_{0.5}$BiS$_2$. The exchange-correlation potential was treated using the generalized gradient approximation (GGA) based on the Perdew-Burke-Ernzerhof (PBE) approach \cite{Perdew_PRL77}. The spin-orbit coupling (SOC) was included as a second variational step self-consistently. The radii of the muffin-tin sphere $R_{MT}$ were 2.30 Bohr for Nd, 2.36 Bohr for Bi, 2.36 Bohr for S and 2.04 Bohr for O. A $17\times17\times5$ $k$-point mesh has been utilized in the self-consistent calculations. The truncation of the modulus of the reciprocal lattice vector $K_{max}$, which was used for the expansion of the wave functions in the interstitial region, was set to $R_{MT}\times K_{max} = 7$.


\begin{figure}[!t]
\begin{center}
\includegraphics[width=3.4 in]{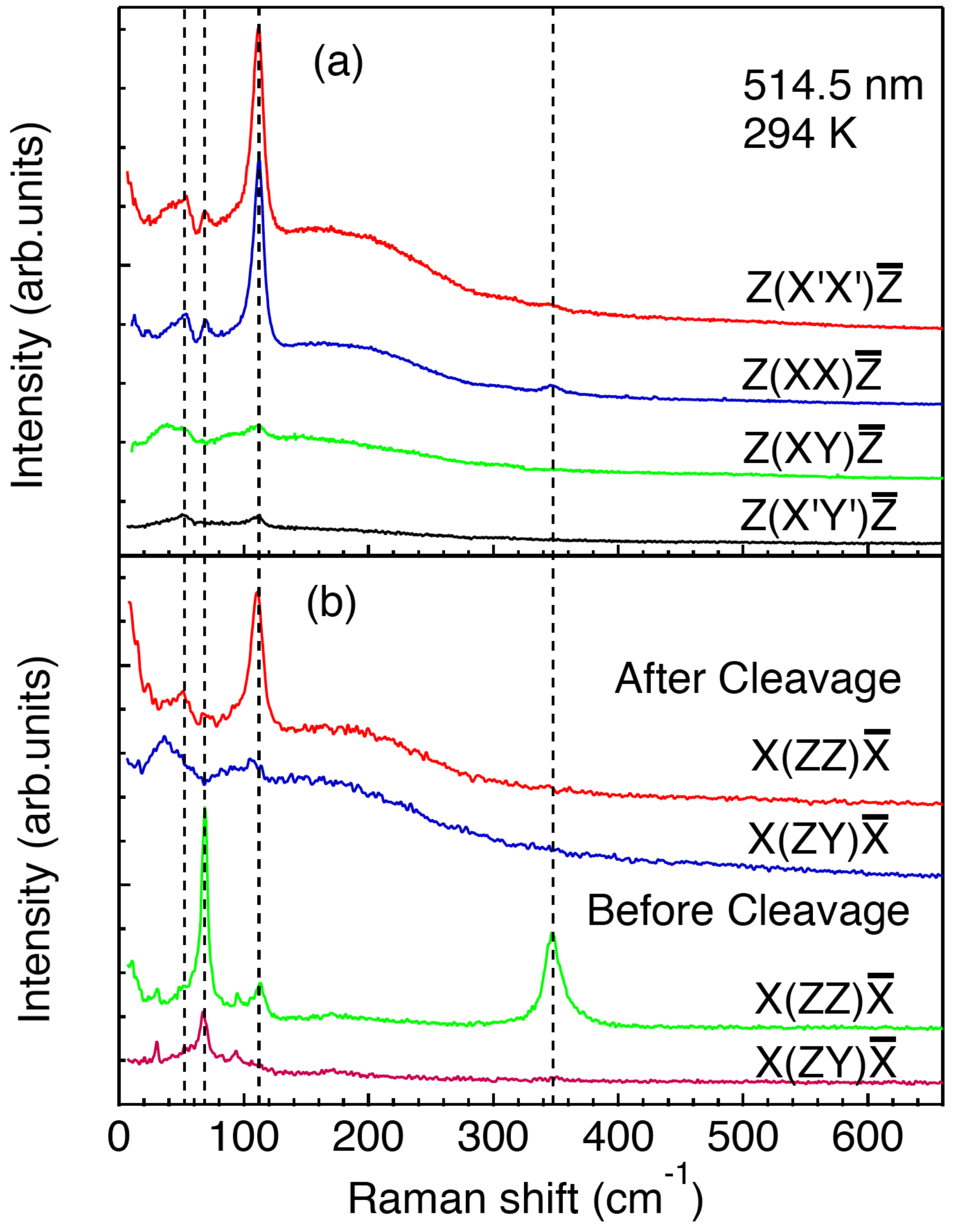}
\caption{\label{phonons}(Color online). (a) Polarization dependence of the $ab$ plane Raman spectra of Nd(O,F)BiS$_2$ recorded at 294 K with a 514.5 nm laser excitation. (b) Polarization dependence of the $ac$ plane spectra at 294 K. The dashed lines in (a) and (b) are guides to the eye for the phonon peak positions.}
\end{center}
\end{figure}

In Fig. \ref{phonons}(a) and \ref{phonons}(b), we show the Raman spectra of Nd(O,F)BiS$_2$ recorded at room temperature under various polarization configurations in the $ab$ plane and in the $ac$ plane, respectively. The symmetry of the modes observed are determined by the Raman tensors corresponding to the $D_{4h}$ symmetry group, which are expressed as:

\begin{displaymath}
\textrm{A$_{1g}$=}
\left(\begin{array}{ccc}
a & 0 &0\\
0 & a &0\\
0 & 0 &b
\end{array}\right)
, \textrm{B$_{1g}$=}
\left(\begin{array}{ccc}
c & 0 &0\\
0 & -c &0\\
0 & 0 &0\\
\end{array}\right),
\end{displaymath}
\begin{displaymath}
\textrm{B$_{2g}$=}
\left(\begin{array}{ccc}
0 & d &0\\
d & 0 &0\\
0 & 0 &0\\
\end{array}\right),
\textrm{E$_{g}$=}
\left\{\left(\begin{array}{ccc}
0 & 0 &0\\
0 & 0 &e\\
0 & e &0\\
\end{array}\right),
\left(\begin{array}{ccc}
0 & 0 &-e\\
0 & 0 &0\\
-e & 0 &0\\
\end{array}\right)
\right\}.
\end{displaymath}

For perfectly aligned crystals, pure A$_{1g}$ symmetry is obtained in the $x(zz)\bar{x}$ configuration. In this channel, we detect a sharp peak at 112.4 cm$^{-1}$ and a weak peak at 52.6 cm$^{-1}$. As summarized in Table \ref{EXP_LDA}, they correspond to $c$-axis vibrations of out-of-plane S atoms and Bi atoms, respectively. In Fig.\ref{phonons}(a), we also show that these two A$_{1g}$ peaks survive in the $z(xx)\bar{z}$ and $z(x'x')\bar{z}$ configurations, for which the A$_{1g}$ signal is mixed with signals from the B$_{1g}$ and B$_{2g}$ channels, respectively. In the $x(zy)\bar{x}$ E$_g$ configuration, we observe a weak peak at 36.5 cm$^{-1}$, which is also seen in the $z(xy)\bar{z}$ B$_{2g}$ channel. Although the origin of this mode remains uncertain, it may relate to a Nd$^{3+}$ crystal-field excitation, the local $C_4$ symmetry of the Nd$^{3+}$ ions implying different selection rules for such transitions. We also observe a broad hump at 200 cm$^{-1}$ in the four in-plane polarization configurations as well as in the $ac$ plane measurements, that we tentatively associate to impurities or inhomogeneity. In addition, two phonon peaks at 68.6 cm$^{-1}$ and 346.7 cm$^{-1}$ in the $z(xx)\bar{z}$ and $z(x'x')\bar{z}$ configurations are observed in samples before cleavage and are attributed to an impurity phase at the surface of the samples.

\begin{figure}[!t]
\begin{center}
\includegraphics[width=3.4 in]{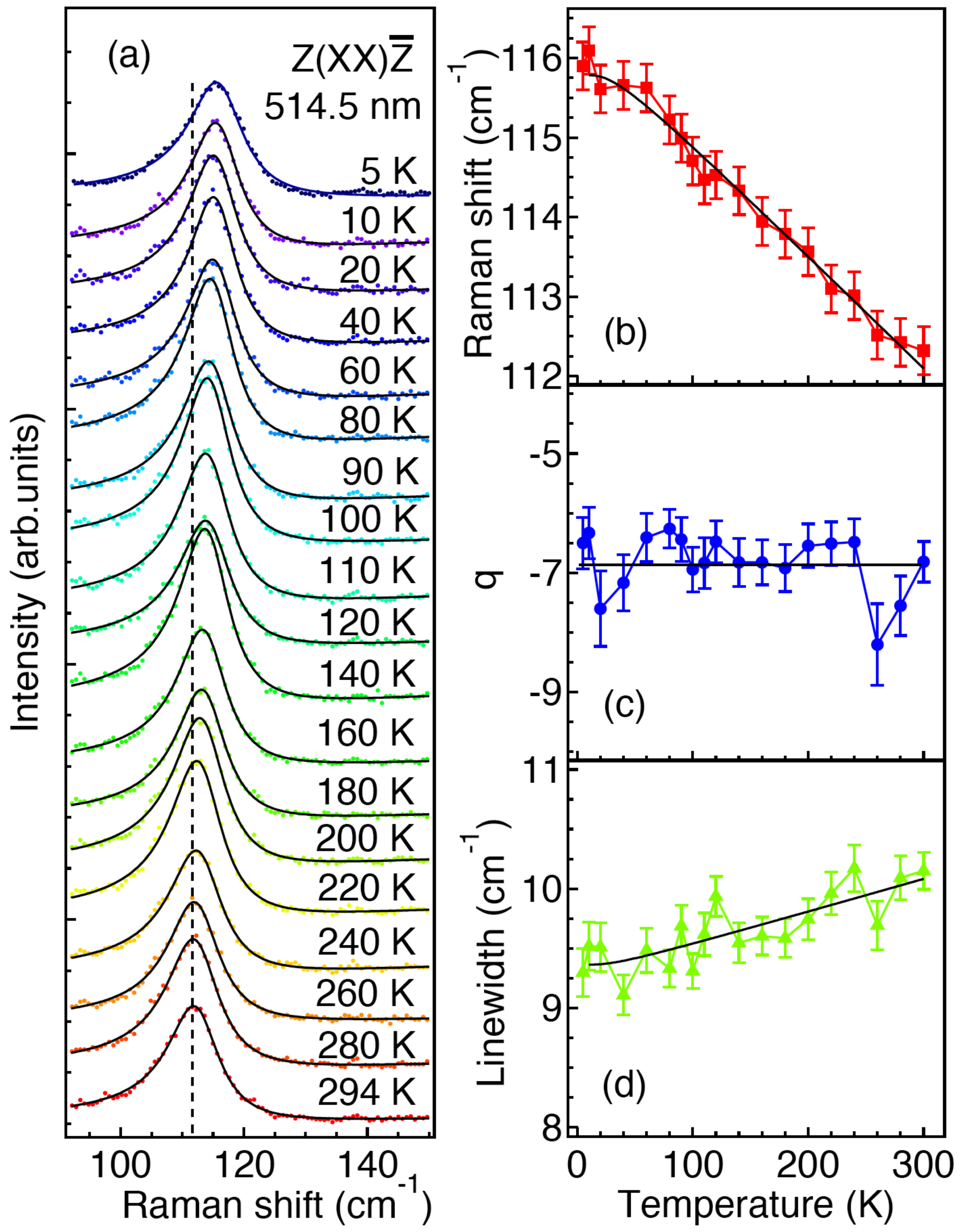}
\end{center}
\caption{\label{T_dependent}(Color online). (a) Temperature dependence of the Raman spectra for the A$_{1g}$ mode at 112.4 cm$^{-1}$. (b)-(d) Peak position , Fano asymmetry factor $q$ and linewidth (full-width-at-half-maximum), respectively, of the A$_{1g}$ mode at 112.4 cm$^{-1}$.  The curves in (b) and (d) are fits to the Eqs. \eqref{eq_omega} and \eqref{eq_gamma}. The line in (c) is the averaged Fano asymmetry factor $q$.}
\end{figure}

To investigate the possible role of the electron-phonon coupling in the SC transition, we cooled the samples down to 5 K. In Fig. \ref{T_dependent}(a), we display the temperature dependence of the A$_{1g}$ peaks at 112.4 cm$^{-1}$. As expected, this peaks become a little harder (3.8 cm$^{-1}$) and sharper (0.64 cm$^{-1}$) with temperature decreasing. A remarkable feature is an asymmetric line shape at all temperatures that suggests an electron-phonon coupling in this system. We found that the Breit-Wigner-Fano (BWF) \cite{Klein.M.V} resonance line shape (in short Fano resonance line shape) can fit the A$_{1g}$ peak 112.4 cm$^{-1}$  better than the Lorentzian or Gaussian line shape. The Fano resonance effect originates from a quantum interference between a discrete state (in this case a phonon mode) and an electron continuum, in which the excited eigenstates are a mixture of the discrete and continuum states \cite{Fano_PhysRev124}. The Fano line shape is described by the equation:

\begin{equation}
I(\omega)=A\frac{[q+2(\omega-\omega_0)/\Gamma]^2}{1+[4(\omega-\omega_0)/\Gamma]^2},
\end{equation}

\noindent where $A$ is the intensity, $\omega_0$ is the renormalized phonon frequency in the presence of the coupling, $q$ is the asymmetry parameter, and $\Gamma$ is the linewidth parameter, which is related to the phonon lifetime. The peak maximum is located at $\omega_{max}=\omega_0+\Gamma/2q$ and its full-width at half-maximum (FWHM) is $\Gamma(q^2+1)/|q^2-1|$. Usually, $1/q$ is referred to as the coupling strength. In the limit of $|1/q|\rightarrow0$, the Fano resonance line shape reduces to the Lorentzian line shape.

In Fig. \ref{T_dependent}(a), we display the BWF fits to the Raman spectra at different temperatures ranging from 5 K to 294 K. In Figs. \ref{T_dependent}(b)-\ref{T_dependent}(d), we display the fit results of the peak position, asymmetry parameter $q$ and the linewidth of the A$_{1g}$ peak at 112.4 cm$^{-1}$, respectively. The peak position and linewidth can be fit well by Eqs. \eqref{eq_omega} and \eqref{eq_gamma}, which describe the temperature dependences derived from a model with anharmonic phonons decaying into acoustic phonons with the same frequencies and opposite momenta \cite{Klemens_PhysRev148,Menendez_PRB29}:

\begin{equation}
\label{eq_omega}
\omega_{ph}(T)=\omega_{0}-C\left( 1+\frac{2}{e^{\frac{\hbar\omega_0 }{ 2k_{B}T}} -1} \right )
\end{equation}
\begin{equation}
\label{eq_gamma}
\Gamma_{ph}(T)=\Gamma_{0}+\Gamma\left( 1+\frac{2}{e^{\frac{\hbar\omega_0 }{ 2k_{B}T}} -1} \right ),
\end{equation}

\noindent where $k_B$ is the Boltzmann constant, $C$ and $\Gamma$ are positive constants, $\omega_0$ is the bare phonon frequency, and $\Gamma_0$ is a residual, temperature-independent linewidth. From the fits, we extract $\omega_0=116.4$ cm$^{-1}$, $C=0.5593$ cm$^{-1}$, $\Gamma=9.23$ cm$^{-1}$ and $\Gamma_0=0.115$ cm$^{-1}$. The asymmetry parameter $q$ recorded in this configuration is nearly independent of temperature and averages to -7.

\begin{figure}[!t]
\begin{center}
\includegraphics[width=3.4 in]{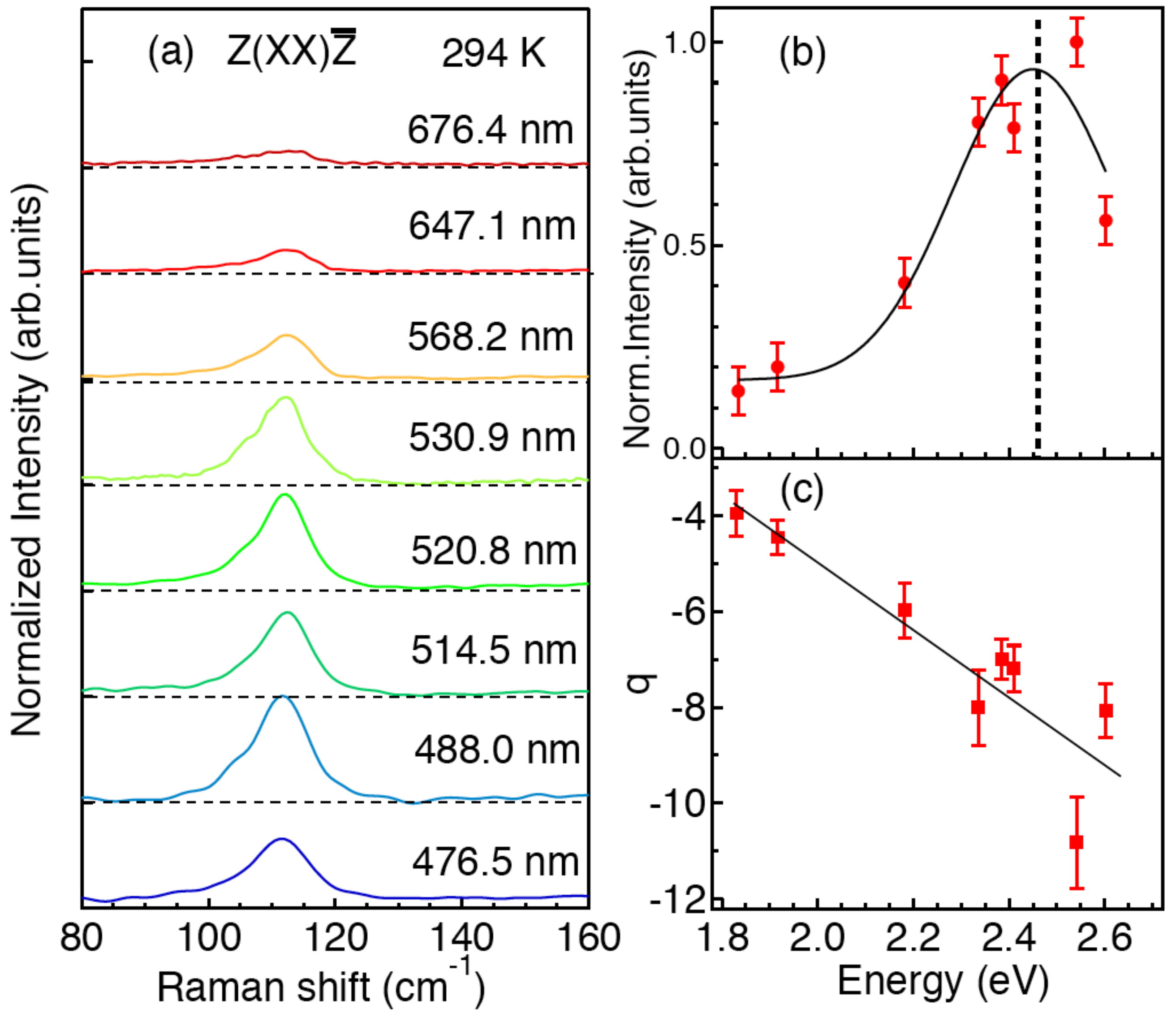}
\end{center}
\caption{\label{laser_dependent}(Color online). (a) Photon energy dependence of the Raman spectra for the A$_{1g}$ mode at 112.4 cm$^{-1}$. (b) Intensity of the A$_{1g}$ mode at 112.4 cm$^{-1}$ with different incident laser normalized by the intensity at 488.0 nm. The black curve is a Gaussian fit to the data. (c) Fano asymmetry factor $q$ of the A$_{1g}$ mode at 112.4 cm$^{-1}$ at different incident laser lines. The curve in black is a linear fit of the data point.}
\end{figure}

We also studied the 112.4 cm$^{-1}$ mode using different incident laser energies, as shown in Fig. \ref{laser_dependent}(a). We observe a strong dependence of the Raman intensity as a function of the laser excitation, as illustrated in Fig. \ref{laser_dependent}(b). More precisely, a maximum of intensity is observed for an energy of 2.45 eV (506 nm), as deduced from a Gaussian fit. The intensity at the maximum is about 6 times higher than for a 1.8 eV laser excitation. Interestingly, this resonance energy is quite similar to the plasma frequency determined from optical measurements on the same batch of samples \cite{Wang_arxiv2014}. Based on the 7\% doping electron per Bi atom deduced from ARPES data \cite{Zeng.arXiv1402.1833} also recorded on samples from the same batch, the optical spectra were interpreted in terms of several interband transitions, including one at about 2.1 eV related to O and Bi $p$ bands. In any case, the resonance observed here for our Raman data likely has an electronic origin, reinforcing the assumption of an electron-phonon coupling. We also found that the asymmetry parameter $q$ decreases almost linearly with the excitantion energy increasing, as shown in Fig. \ref{laser_dependent}(c).

The electron-phonon coupling constant $\lambda_i$ associated to a particular mode $i$ can be estimated from the Allen formula \cite{Allen_PRB12,Allen_PhysRevB.6.2577}, which relates the linewidth $\gamma_i$ of the phonon mode $i$ to the dimensionless electron-phonon coupling constant \cite{Varma1991,Rodriguez_PRB42,Zhou_PRB48}:

\begin{equation}
\gamma_i=\frac{1}{g_i}\pi N(\epsilon_{f})\lambda_i \omega_{bi}^2
\end{equation}

\noindent where $N(\epsilon_{f})$ is the electronic density of states at the Fermi level per eV per spin per unit cell, $g_i$ is the mode degeneracy and $\omega_{bi}$ is the bare frequency in the absence of electron-phonon interaction. In our Raman spectra, the A$_{1g}$ mode at 112.4 cm$^{-1}$ shows the largest asymmetry line shape, which is relevant in the context of conventional superconductivity. Since the peak position of this mode varies only slightly with temperature, we take the bare frequency $\omega_{bi}$ as 112.4 cm$^{-1}$. With $g_i=1$ for a A$_{1g}$ mode, $\gamma_i=9.4$ cm$^{-1}$ as deduced from Fig. \ref{T_dependent}(d) and $N(\epsilon_{f}) = 2.8178$ states/eV/spin/unit cell from our band structure calculations of NdO$_{0.5}$F$_{0.5}$BiS$_2$, we get $\lambda_i = 0.68$, a value close to $\lambda = 0.85$ estimated from calculations \cite{Wan_PRB87,Li_EPL101}. We caution though that several factors may affect this value in one way or the other. For example, ARPES results \cite{Zeng.arXiv1402.1833,Ye.arXiv1402.2860} suggest that the actual doping level might be smaller than that inferred from the nominal concentration, which would lead to a larger coupling strength. In contrast, our analysis may underestimate the contribution of impurities to the linewidth, which would result in a smaller coupling strenght.

Within the BCS framework, $T_c$ can be estimated \cite{Zhou_PRB48,Kortus_PRL86,Shukla_PRL90} from the experimental $\lambda$ and the McMilan relation \cite{McMillan_PhysRev167} modified by Allen and Dynes \cite{Allen_PRB12}:

\begin{equation}
T_c=\frac{\hbar \omega_{log}}{1.2k_B}\exp\left [-\frac{1.04(1+\lambda)}{\lambda-\mu^*(1+0.62\lambda)}\right ],
\end{equation}

\noindent where $\omega_{log}$ is the logarithmic average of all the phonon frequencies, $\lambda=\sum_i\lambda_i$ is the electron-phonon coupling constant and $\mu^*$ is the unretarded screened Coulomb interaction. Unfortunately, here our experimental knowledge of the phonon frequencies and their contributions to the electron-phonon coupling constant is very limited. Limiting ourselves to the mode at 112.4 cm$^{-1}$ and using $\mu^*=0.1$ as commonly used for similar materials \cite{Kortus_PRL86_MgB2}, we found $T_c=5.2$ K, which is similar to the experimental value of 4 K. Although our approximations remain quite rough, our experimental observation of an electron-phonon coupling with a strength compatible with the one expected for a phonon-driven pairing mechanism suggests that Nd(O,F)BiS$_2$ is possibly a conventional superconductor, in agreement with a penetration depth that follows an exponential temperature dependence well below $T_c$ \cite{Jiao_arXiv2014}.

In summary, we have performed polarized Raman scattering measurements on the newly discovered superconductor Nd(O,F)BiS$_2$ ($T_c = 4$ K). We observe 2 Raman active  modes, with frequencies in decent agreement with first-principles calculations. The phonon spectra reveal a photon energy resonance at 2.45 eV and a Fano line shape suggestive of an electron-phonon coupling. Our results suggest that the BiS$_2$ based superconductors are possibly phonon-mediated BCS superconductors.

This work was supported by grants from CAS (2010Y1JB6), MOST (2010CB923000,  2011CBA001000, 2011CBA00101, 2012CB821403 and 2013CB921703) and NSFC (11004232, 11034011/A0402, 11234014 , 11274362 and 11274356) from China.

\bibliography{biblio_long}

\end{document}